\documentclass[11pt,round]{article}
\usepackage{setspace}
\usepackage[top=2cm,bottom=2cm, left=2cm, right=2cm]{geometry}
\usepackage{natbib}
\usepackage[colorlinks=true, allcolors=blue]{hyperref}
\usepackage{bm}
\usepackage{graphicx}
\usepackage{natbib}
\usepackage{booktabs}
\usepackage{amsfonts}
\usepackage{amsmath,amssymb,amsthm}
\usepackage{algorithm}
\usepackage{algpseudocode}

\title{Modelling structural zeros in compositional data via a zero--censored multivariate normal model}
\author{
Michail Tsagris$^1$ and Nader Alharbi$^2$  \\
$^1$ Department of Economics \& Institute of Research in Education and Digital Social \\ Sciences and Humanities, University of Crete, Gallos Campus, Rethimno, Greece \\
\href{mailto:mtsagris@uoc.gr}{mtsagris@uoc.gr} \\
$^2$ King Saud bin Abdulaziz University for Health Sciences, \& King Abdullah \\ International Medical Research Center, Riyadh, Saudi Arabia \\
\href{mailto:alharbina@ksau-hs.edu.sa}{alharbina@ksau-hs.edu.sa} 
}

\begin{document}
\maketitle

\begin{center}
{\bf Abstract}
\end{center}
Compositional data are multivariate data constrained to lie within the simplex. When zero values are present, most methods fail to apply unless a zero value replacement is performed prior to the analysis. In this paper we present a new model for analyzing compositional data with zero values present. Unlike the model of \cite{butler2008} that projects every zero value towards a vertex, we project them onto their corresponding edge and assume that zero values are censored values from a multivariate normal distribution. We utilize the EM algorithm to estimate the model parameters, in a computationally efficient manner. Simulated and real data illustrate the performance of the proposed model.  

\section{Introduction}
Compositional data are non--negative multivariate vectors that convey only relative information, typically normalized to sum to one, and their treatment of missing values must account for this restrictive sample space. Specifically, the sample space is the standard simplex, given by
\begin{eqnarray} \label{simplex}
\mathbb{S}^{d}=\left\lbrace(x_1,...,x_D)^\top \bigg\vert x_i \geq 0,\sum_{i=1}^Dx_i=1\right\rbrace,
\end{eqnarray}
where $D$ denotes the number of variables, usually referred to as components, and $d=D-1$.

Compositional data arise across numerous application domains (see \cite{tsagris2020} for a range of examples), and a substantial body of literature has been devoted to methodology for their proper analysis. Structural (or essential) zeros are occasionally encountered in compositional data. The term "structural" refers to values that are genuinely zero, such as the proportion of household expenditure allocated to smoking or alcohol by a family. Rounded zeros, by contrast, arise when a component's true value is very small but is recorded as zero. In geology, for instance, the instrument used to measure elemental composition is subject to a detection limit, and values below this threshold go undetected; such a value may correspond either to the complete absence of the element or to a true value smaller than the instrument's detection limit.

Ever since \cite{ait1982}, the log--ratio approach has been the most widely used framework for compositional data analysis. It involves transforming the data to Euclidean space via a log--ratio transformation, after which standard multivariate techniques can be applied. The nature of the logarithmic transformation, however, gives rise to a mathematical difficulty, since the logarithm of zero is undefined. Consequently, zeros pose a substantial obstacle to this strategy, and considerable attention has been directed toward alternative techniques for handling compositional data containing zeros.

This problem has traditionally been addressed through simple imputation techniques, such as imputation by a small value \citep{ait2003}, substitution of the zero by a fraction of the detection limit \citep{palarea2005}, or imputation via the EM algorithm \citep{palarea2007}. These approaches are appropriate when the zeros in question are rounded, in which case the true value may be lower than the estimated one. \cite{scealy2011a} illustrated the difficulties that can arise when such approaches are adopted: the smaller the imputed value, the larger the magnitude of the resulting log--ratio transformed values. Moreover, when a zero is structural rather than rounded, imputation of any kind is fundamentally inappropriate.

\cite{butler2008} proposed a latent Gaussian model for the treatment of zero values, in which points falling outside the simplex are projected orthogonally onto its vertices. This approach, however, can assign excessive probability mass to the vertices, and as the dimensionality of the simplex increases, identifying the appropriate regions onto which such points should be projected becomes increasingly difficult. Maximum likelihood estimation is similarly complicated under this formulation, although the authors address the resulting estimation difficulties through the use of MCMC methods.

In this paper, we introduce an alternative approach to modelling zero values, motivated by the framework proposed by \citet{butler2008}. Instead of orthogonally projecting points outside the simplex onto its boundary, our approach moves each such point along the line connecting it to the center of the simplex. Similar to the model of \citet{butler2008}, we posit an underlying latent multivariate normal distribution, with only the corresponding compositional data being observed. Under this framework, zero values in the observed compositions arise when realizations of the latent distribution fall outside the simplex. We also propose a goodness--of--fit test for the observed pattern of zeros. A key advantage of our proposed model is that, unlike the approach of \citet{butler2008}, its likelihood remains tractable irrespective of the dimensionality of the composition, enabling efficient estimation using the EM algorithm. The main limitation of the proposed approach is that it can accommodate only one zero value in each compositional vector.

The next section introduces the zero--censored model, followed by simulation studies in Section \ref{sec:sims} and a real data example that illustrates the performance of the model in Section \ref{sec:real}. Finally, conclusions close the paper. 

\section{The zero--censored model}
We fit a multivariate normal distribution to the simplex, consisting of two components: one corresponding to observations that lie in the interior of the simplex and another describing points located on its faces. A key, and potentially restrictive, feature of the proposed model is that it assigns zero probability to the vertices of the simplex. To overcome the unit--sum constraint, we initially employ the $\alpha$--transformation \citep{tsagris2011} with $\alpha=1$. Specifically, this transformation centers the simplex and scales it by $D$, the number of components, after which the Helmert sub--matrix is applied from the left to remove the unit--sum constraint.

\subsection{The $\alpha$--transformation} \label{sec:alfa}
For a composition $\bm{x} \in \mathbb{S}^{d}$, the centered log--ratio (clr) transformation is defined in \cite{ait1983} as 
\begin{eqnarray} \label{clr}
\bm{w}_0(\bm{x})=\left ( \log\left ({\frac{x_1}{\prod_{j=1}^Dx_j^{1/D}}} \right ),\ldots, \log\left ({\frac{x_D}{\prod_{j=1}^Dx_j^{1/D}}} \right ) \right ).
\end{eqnarray}
The obvious drawback of the clr is that it still possesses singularity issues because the sum of the transformed data equals zero. To remove the redundant dimension imposed by this constraint \cite{ilr2003} proposed the isometric log--ratio transformation 
\begin{eqnarray} \label{ilr}
\bm{y}_0(\bm{x})=\bm{H}\bm{w}_0(\bm{x}),
\end{eqnarray}
where $\bm{y}_0(\bm{x})$ is a $D-1$ dimensional vector and $\bm{H}$ is the $d \times D$ Helmert \citep{helm1965} sub--matrix\footnote{This is the Helmert matrix after deletion of the first row. This sub--matrix is a standard orthogonal matrix in shape analysis used to overcome singularity problems. For further information, see \cite{dryden1998,le1999}.}, for which
$\bm{HH}^\top=\bm{I}_{d}, \bm{H1}_D=\bm{0}_d$. Left multiplication by the Helmert sub--matrix maps the clr transformed data  onto $\mathbb{R}^d$, thus, in effect, removing the zero sum constraint. 

\cite{tsagris2011} developed the $\alpha$--transformation as a more general transformation than the ilr transformation. Let 
\begin{eqnarray} 
\label{stayalpha}
\bm{u}_{\alpha}(\bm{x})=\left( \frac{x_1^{\alpha}}{\sum_{j=1}^Dx_j^{\alpha}}, \ldots, \frac{x_D^{\alpha}}{\sum_{j=1}^Dx_j^{\alpha}} \right)^\top
\end{eqnarray}
denote the power transformation for compositional data as defined by \cite{ait2003}, where $\alpha$ can take any value, (the authors suggest the interval $[-1,1]$), but when zero values exist in the data, $\alpha$ can take only strictly positive values. In a manner analogous to (\ref{clr}--\ref{ilr}), first define
\begin{eqnarray} \label{alef}
\bm{w}_{\alpha}(\bm{x})=\frac{D\bm{u}_{\alpha}(\bm{x})-1}{\alpha}.
\end{eqnarray}
The $\alpha$--transformation is defined as
\begin{eqnarray} \label{alpha}
\bm{y}_{\alpha}(\bm{x})=\bm{H}\bm{w}_{\alpha}(\bm{x}).
\end{eqnarray}
The transformation in (\ref{alpha}) is a one--to--one transformation which maps data inside the simplex onto a subset of $\mathbb{R}^{d}$ and vice versa for $\alpha \neq 0$. From (\ref{stayalpha}) and (\ref{alef}), we can observe that when $\alpha=1$, the simplex is linearly expanded as the values of the components are simply multiplied by a scalar and then centered. The corresponding sample space of (\ref{alpha}) when $\alpha=1$ is 
\begin{eqnarray} \label{Ad}
\mathbb{A}_1^{D-1}=\left\lbrace{\bf H}{\bf w}_1(\mathbf{x}) \bigg | -1 \leq w_{i,1} \leq D-1,\sum_{i=1}^Dw_{i, 1}=0 \right\rbrace.
\end{eqnarray}

\subsection{The mechanism of the model}
The composition $\bm{x}\in \mathbb{S}^d$ is modelled through a latent Gaussian vector in the unconstrained space obtained via the $\alpha=1$ transformation, before the simplex constraint is imposed. The inverse of the $\alpha$--transformation does not guarantee that each point will lie in the simplex space. The interior points are treated as ordinary continuous observations from a multivariate normal contributing the usual density $g(\bm{y}_i)$ to the likelihood; but whenever a latent draw $\bm{y}_i$ would map to a composition with a negative $j$-th component, we do not observe that value, but instead observe $x_j=0$, i.e. that the event $\{\bm{e}_j^\top \bm{y}_i > c_j\}$ occurred, for the direction $\bm{e}_j$ orthogonal to facet $j$ and threshold $c_j$ determined purely by the simplex geometry. This is a genuine censoring mechanism, the latent value truly exceeded the boundary, and the recorded zero encodes that, so its likelihood contribution is the tail probability $P(\bm{e}_j^\top \bm{Y}_i>c_j)$ rather than the density evaluated at that point.
 
\subsection{The joint density of observed and censored coordinates}
Assume $\bm{Y}_{obs}$ is the part that remains fully observed and $Y_e$ is the part that is censored (known only to exceed $c$). Since this decomposition is an orthogonal change of variables, the joint density of the latent vector, written in these coordinates, is
\begin{equation*}
f_{\bm{Y}}(\bm{y}) = f(y_e, \bm{y}_{obs}),
\end{equation*} 
with no Jacobian correction needed. By the definition of a conditional density, an identity that holds for any joint distribution, not only the normal distribution, the above density factors as the product of the marginal density of the observed part and the conditional density of the censored part given the observed part
\begin{equation*}
f(y_e,y_{obs}) = f_{obs}(y_{obs})\cdot f_{e\mid{obs}}(y_e\mid y_{obs}),
\end{equation*}.

The likelihood contribution of this compositional observation is the probability of the observed event $\{Y_e>c,\ \bm{Y}_{obs}=\bm{y}_{obs}\}$, obtained by integrating the joint density over the unobserved region and evaluating at the observed value of $\bm{Y}_{obs}$:
\begin{equation*}
L = \int_c^\infty f(y_e,\bm{y}_{obs})\,dy_e
 = f_{obs}(\bm{y}_{obs})\int_c^\infty f_{e\mid obs}(y_e\mid \bm{y}_{obs})\,dy_e
 = f_{obs}(\bm{y}_{obs})\cdot\big[1-F_{e\mid obs}(c\mid \bm{y}_{obs})\big],
\end{equation*}

\subsection{The log--likelihood of the model}
Similarly to \citet{butler2008} we can write the log--likelihood as
\begin{eqnarray} \label{zerolog1}
\ell=\sum_{i=1}^{n_1}\log{g\left(\bm{y}_{1i}\right)}+\sum_{i=1}^n\log{\left|J\right|}+
\sum_{i=1}^{n_2}\log{\int_{k_j}^{\infty}f_i\left(\bm{y}_{1i}\right)dy_j},  
\end{eqnarray}
where $\bm{y}_1$ is the $\alpha$--transformation \citep{tsagris2011} with $\alpha=1$, $g\left(.\right)$ is the density of the multivariate normal for the data lying inside the simplex, $f_i\left(.\right)$ is the density of the $i$--th point lying outside the simplex given that it is in a line going through the origin. The $n_1$ is the number of points lying inside the interior of the simplex and $n_2$ denotes the number of points on the faces of the simplex. The line integral refers to the $i$--th observation lying on the face the simplex for which the integral is calculated along the $j$--th component, with $j \in [1,...,D]$, where $D$ is the number of components. Finally, $\left|J\right|$ is the Jacobian determinant of the $\alpha$--transformation with $\alpha=1$ \citep{tsagris2020}.  

\begin{figure}[!h]
\includegraphics[scale=0.9,trim=0 0 0 0]{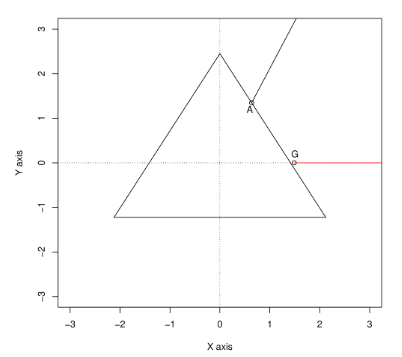}
\centering
\caption{Ternary diagram showing the zero projection. We want to evaluate the line integral of the multivariate normal distribution from A to $\infty$ along the black line. For this reason we rotate the point onto the X axis and find the integral from G to $\infty$.} 
\label{zero}
\end{figure}

We will need to  calculate the line integral of this component in the multivariate density from that point to infinity. Therefore, the log--likelihood consists of the density inside the interior of the simplex and the density on the faces, thus assigning zero probability on the vertices (and to the edges when $D>3$). We will express (\ref{zerolog1}) in a more convenient way, by dropping the subscript $1$ from $\bm{y}$ for convenience purposes
\begin{eqnarray} \label{zerolog2}
\ell = -\frac{n_1}{2}\log{\left|2 \pi \bm{\Sigma} \right|}-
\frac{1}{2}\sum_{i=1}^{n_1}\left(\bm{y}_i-\bm{\mu}\right)^T\bm{\Sigma}^{-1}\left(\bm{y}_i-\bm{\mu}\right) +\sum_{i=1}^{n_2}\log{\int_{c_1}^{\infty}f_i\left(\bm{z}\right)dz_1 } +
\left(nd+\frac{n}{2}\right)\log{D},
\end{eqnarray}
where $n=n_1+n_2$ is the full sample size. The vector inside the integral has changed from $\bm{y}_i$ to $\bm{z}_i$ with $\bm{z}=\bm{Oy}$, where $c_1$ and $\bm{O}$ will be explained below in the Gram--Schmidt orthonormalization process \citep{strang1988}. In order to calculate the line integral we will first perform a rotation towards one arbitrary direction. For convenience purposes, we chose the first direction, the $X$ axis for instance in the two dimensions as seen in Figure \ref{zero}. The rotation takes place by multiplying the vector with the zero component (after the $\alpha$--transformation) by an orthonormal matrix. 

Figure \ref{zero} shows a graphical example of the rotation in $\mathbb{R}^2$. The red line integral is calculated through a normal distribution whose parameters are rotated via the Gram--Schmidt orthonormalization process in the same way the black line was rotated to the red line. This is one example of a composition with a zero value in one of its components.

\subsection{Gram--Schmidt orthonormalization process}
The Gram--Schmidt orthonormalization process is described as follows. Suppose we have a vector $\bm{v}$ in $\mathbb{R}^d$ and we want to rotate it to the line defined by the unit vector $\bm{e}_1=\left(1,0,\ldots,0\right)^\top$. We have to find an orthonormal basis first using the Gram--Schmidt orthonormalization process. Let us denote the projection operation of a vector $\bm{v}$ onto $\bm{u}$ by 
\begin{eqnarray*}
proj_{\bm{u}}\left(\bm{v}\right)=\frac{\langle \bm{v},\bm{u} \rangle}{\langle \bm{u},\bm{u} \rangle}\bm{u}.
\end{eqnarray*}
Then the following operations take place
\begin{eqnarray*}
\begin{array}{ccccc}
\bm{u}_1 = & \bm{v}_1 &                                         &  &   \text{and} \ \ \bm{o}_1 = \frac{\bm{u}_1}{ \Vert \bm{u}_1 \Vert }, \\
\bm{u}_2 = & \bm{v}_2 & -proj_{\bm{u}_1}\left(\bm{v}_2\right) &  &   \text{and} \ \ \bm{o}_2=\frac{\bm{u}_2}{ \Vert \bm{u}_2 \Vert }, \\
\bm{u}_3 = & \bm{v}_3 & -proj_{\bm{u}_1}\left( \bm{v}_3\right) & -proj_{\bm{ u}_2}\left(\bm{v}_3\right) &   \text{and} \ \ \bm{o}_3=\frac{\bm{u}_3}{ \Vert \bm{u}_3 \Vert }, \\
\vdots      &           &                                         &   &  \vdots      \\
\bm{u}_d = & \bm{v}_d & -\sum_{i=1}^{d-1}proj_{\bm{u}_i}\left(\bm{v}_d\right) & &  \text{and} \ \ \bm{o}_d=\frac{ \bm{u}_d}{ \Vert \bm{u}_d \Vert }, \\
\end{array}
\end{eqnarray*}
where $\langle \cdot \rangle $ denotes the inner product of two vectors. Denote the matrix of the orthonormalized vectors $\bm{o}$ by  $\bm{O}=\left[\bm{o}_1, \ldots, \bm{o}_d \right]$. Then all we have to do to get $\bm{c}$ is $\bm{c}=\bm{Oe}_1$ and the first element of the vector $\bm{c}$ is the term $c_1$ we saw in (\ref{zerolog2}).

\subsection{The log--likelihood of the zero--censored model}
Since the integral of (\ref{zerolog2}) is with respect to the first variable of the multivariate normal distribution, we can use the conditional normal distribution to write (\ref{zerolog2}) in a more attractive form as
\begin{eqnarray} \label{zerolog3}
\ell &=& -\frac{n_1}{2}\log{\left|2 \pi \bm{\Sigma} \right|}-
\frac{1}{2}\sum_{i=1}^{n_1}\left(\bm{y}_i-\bm{\mu}\right)^T\bm{\Sigma}^{-1}\left(\bm{y}_i-\bm{\mu}\right)+ \left(nd+\frac{n}{2}\right)\log{D}  \nonumber \\
& & +\sum_{i=1}^{n_2}\log{\left[f_i\left(\bm{z}_{-1};\bm{\mu}_{iz^*},\bm{\Sigma}_{iz^*}\right)
\int_{c_{1i}}^{\infty}f_i\left(z_{1i} \vert \bm{ z}_{-1};\mu_{i,con},\sigma^2_{i,con}\right)dz_{1i}\right]},
\end{eqnarray}
where $\bm{z}_{-1}$ means all elements except from the first one. $f_i\left( \bm{z}_{-1};\bm{\mu}_z^*,\bm{\Sigma}_z^*\right)$ is the density of the multivariate normal with parameters $\left(\bm{\mu}_z^*,\bm{\Sigma}_z^*\right)$ calculated at $\bm{z}_{-1}$ and $f_i\left(z_{1i} \vert \bm{z}_{-1};\mu_{i,con},\sigma^2_{i,con}\right)$ is the density of the conditional univariate normal with parameters $\left(\mu_{i,con},\sigma^2_{i,con}\right)$ calculated at $z_{1i}$. The conditional distribution of a multivariate normal is still a normal \citep{mardia1979} and the following relationships hold true   
\begin{eqnarray*}
\left( \bm{X}_1,\bm{X}_2 \right) \sim N_d\left(\left(\bm{\mu}_1,\bm{\mu}_2\right)^T,\bm{\Sigma}\right), \ \ \text{then} \ \
\bm{X}_1 \big\vert \bm{X}_2 \sim N_d\left(E\left(\bm{X}_1 \big\vert \bm{X}_2 \right),V\left( \bm{X}_1 \big\vert \bm{X}_2\right) \right),
\end{eqnarray*}
where 
\begin{eqnarray*}
E\left(\bm{X}_1 \big\vert \bm{X}_2 \right) =\bm{\mu}_1+\bm{\Sigma}_{12}\bm{\Sigma}_{22}^{-1}\left(\bm{X}_2-\bm{\mu}_2\right) \ \ \text{and} \ \ V\left(\bm{X}_1 \big\vert \bm{X}_2 \right) =\bm{\Sigma}_{11}-\bm{\Sigma}_{12}\bm{\Sigma}_{22}^{-1}\bm{\Sigma}_{21}.
\end{eqnarray*}
Hence, using these relationships we can calculate the parameters of the normal density appearing inside the integral of (\ref{zerolog3}). Thus, we have the following relationships
\begin{eqnarray*}
\bm{\mu}_{iz}=\bm{B}_i\bm{\mu} \ \ \text{and} \ \ \bm{\Sigma}_{iz}=\bm{B}_i\bm{\Sigma B}_i^T,
\end{eqnarray*}
where the index $i$ is used to indicate the $i$--th observation and $\bm{B}_i$ means the rotation matrix (calculated from the Gram--Schmidt orthonormalization process) for the $i$--th observation. The rotation matrix rotates the vector $\bm{y}_i$ to the line defined by the unit vector $\bm{e}_1$. 
Now,
\begin{eqnarray*}
\bm{\mu}_{iz^*}=\bm{\mu}_{iz,-1} \ \ \text{and} \ \ \bm{\Sigma}_{iz^*}=\bm{\Sigma}_{iz}[-1,-1]
\end{eqnarray*}
and
\begin{eqnarray*}
\mu_{i,con} &=& \bm{\mu}_{iz,1}-\bm{\Sigma}_{iz}[1,]\bm{\Sigma}_{iz}[-1,-1]^{-1}\bm{\mu}_{iz,-1} \\ 
\text{and}  \ \ \sigma^2_{i,con} &=& \bm{\Sigma}_{iz}[1,1]-\bm{\Sigma}_{iz}[1,]\bm{\Sigma}_{iz}[-1,-1]^{-1}\bm{\Sigma}_{iz}[1,]^T,
\end{eqnarray*}
where $\bm{\mu}_{iz,-1}$ and $\bm{\Sigma}[-1,-1]$ denote the mean vector $\bm{\mu}_{iz}$ and the covariance matrix $\bm{\Sigma}$, respectively without the first element and $\bm{\Sigma}[1,]$ indicates the first row of $\bm{\Sigma}$. 

The rationale is to multiply each vector by the rotation matrix $\bm{B}_i$ and rotate the data onto the first axis, thus the new vector is denoted by $\bm{c}_i=\left(c_{1i},0\ldots,0\right)$. Thus (\ref{zerolog3}) can be written as 
\begin{eqnarray} \label{zerolog4}
\ell &=& -\frac{n_1}{2}\log{\left|2 \pi \bm{\Sigma} \right|}-\frac{1}{2}\sum_{i=1}^{n_1}\left(\bm{y}_i-\bm{\mu}\right)^T\bm{\Sigma}^{-1}\left(\bm{y}_i-\bm{\mu}\right) + \left(nd+\frac{n}{2}\right)\log{D}  \nonumber \\
& & + \sum_{i=1}^{n_2}\log{f_i\left(\bm{0};\bm{\mu}_{iz^*},\bm{\Sigma}_{iz^*}\right)}+\sum_{i=1}^{n_2}\log{\left[1-\Phi\left(\frac{c_{1i}-\mu_{i,con}}{\sigma_{i,con}}\right)\right]},
\end{eqnarray}
where $\Phi\left(.\right)$ is the cumulative distribution of a standard normal random variable. The final form of the log--likelihood (\ref{zerolog4}) is the form maximized numerically. The index $i$ in each of the parameters (for the compositions which contained one zero value) indicates that each composition with one zero value had to be projected onto the face and thus its contribution to the parameters is different. 

\subsection{MLE using the EM algorithm}
We only know $z_{1i} > c_{1i}$, and treat for each censored observation, the true (unobserved) rotated coordinate $z_{1i}$ as latent data. Given this, $(z_{1i}, \bm{z}_{-1,i}=\bm{0})$ would constitute a complete observation, from which $\bm{y}_i = \bm{B}_i^\top \bm{z}_i$ could be recovered, and the complete--data log--likelihood would be an ordinary multivariate normal log--likelihood over all $n$ points.
 
The complete--data log--likelihood, as a function of $(\bm{\mu},\bm{\Sigma})$, is
\begin{eqnarray*}
\ell_c(\bm{\mu},\bm{\Sigma}) = -\frac{n}{2}\log|2\pi\bm{\Sigma}| 
-\frac{1}{2}\sum_{i=1}^n (\bm{y}_i - \bm{\mu})^\top \bm{\Sigma}^{-1} (y_i-\bm{\mu}),
\end{eqnarray*}
where for censored observations $\bm{y}_i$ is the (unobserved) completed vector. 

Since this is linear in the sufficient statistics $\bm{y}_i$ and $\bm{y}_i\bm{y}_i^\top$, the E--step reduces to computing their conditional expectations given the observed data and current parameter estimates. For the E--step we separate between the interior points (vectors within the simplex) and the censored points. For the interior points there is nothing to impute, hence
\begin{eqnarray*}
E[\bm{y}_i] = \bm{y}_i, \ \ E[\bm{y}_i\bm{y}_i^\top] = \bm{y}_i\bm{y}_i^\top, \ \ i=1,\dots,n_1.
\end{eqnarray*}
 
For the censored points only $z_{1i}$ is missing, and it is known to exceed $c_{1i}$. Define the standardized threshold
\begin{eqnarray*}
\alpha_i = \frac{c_{1i}-\mu_{i,con}^{(t)}}{\sigma_{i,con}^{(t)}},
\end{eqnarray*}
using the current--iteration values $\mu_{i,con}^{(t)},\sigma_{i,con}^{(t)}$ (computed from $\bm{\mu}^{(t)},\bm{\Sigma}^{(t)}$ via $\bm{B}_i$, as above). The moments of a normal distribution, truncated to $(c_{1i},\infty)$ yield
\begin{align}
\hat z_{1i} &= E[z_{1i}\mid z_{1i}>c_{1i}] = \mu_{i,con}^{(t)} + \sigma_{i,con}^{(t)}\,\lambda(\alpha_i),
\label{eq:trunc-mean}\\
E[z_{1i}^2 \mid z_{1i}>c_{1i}] &= \big(\sigma_{i,con}^{(t)}\big)^2\Big(1-\lambda(\alpha_i)\big(\lambda(\alpha_i)-\alpha_i\big)\Big) + \hat z_{1i}^2,
\label{eq:trunc-second-moment}
\end{align}
where $\lambda(\alpha) = \phi(\alpha)/[1-\Phi(\alpha)]$ is the inverse
Mills ratio.
 
Since $\bm{B}_i$ is driven by the observation's own direction, $z_{-1,i}=0$, so the rotated vector is $\hat{\bm{z}}_i = (\hat z_{1i}, 0,\dots,0)$, and rotating back using $\bm{B}_i^\top$'s first column yields
\begin{eqnarray*}
E[\bm{y}_i] = \frac{\hat z_{1i}}{a_i}\, \bm{y}^*_i,
\qquad
E[\bm{y}_i\bm{y}_i^\top] = \frac{E[z_{1i}^2\mid z_{1i}>c_{1i}]}{a_i^2}\, \bm{y}^*_i\bm{y}_i^{*\top},
\end{eqnarray*}
where $\bm{y}^*_i$ denotes the vector with a censored point. (The second identity uses $E[\bm{z}_i\bm{z}_i^\top] = \mathrm{diag}(E[z_{1i}^2], 0,\dots,0)$
together with $\bm{y}_i = \bm{B}_i^\top \bm{z}_i$, so $E[\bm{y}_i\bm{y}_i^\top] = \bm{B}_i^\top E[\bm{z}_i\bm{z}_i^\top]\bm{B}_i = E[z_{1i}^2]\,(\bm{B}_i^\top \bm{e}_1)(\bm{B}_i^\top \bm{e}_1)^\top = E[z_{1i}^2]\,(\bm{y}^*_i/a_i)(\bm{y}^*_i/a_i)^\top$.)
 
Given the complete sufficient statistics, the M--step is the ordinary
multivariate normal MLE:
\begin{align}
\bm{\mu}^{(t+1)} &= \frac{1}{n}\left(\sum_{i=1}^{n_1} \bm{y}_i + \sum_{i=1}^{n_2} E[\bm{y}_i]\right),
\label{eq:mstep-mu}\\
\bm{\Sigma}^{(t+1)} &= \frac{1}{n}\left(\sum_{i=1}^{n_1} \bm{y}_i\bm{y}_i^\top + \sum_{i=1}^{n_2} E[\bm{y}_iy_i^{\top}]\right) - \bm{\mu}^{(t+1)}\left(\bm{\mu}^{(t+1)}\right)^\top.
\label{eq:mstep-sigma}
\end{align}
 
\subsection{Random values generation}
Let $D\ge2$, $d=D-1$, and let $\bm{h}_j\in\mathbb{R}^d$ denote the $j$--th
column of the Helmert sub--matrix $\bm{H}$. Define the latent--to--composition
map $\bm{A}^{-1}:\mathbb{R}^{d}\to\mathbb{R}^D$ by
\begin{eqnarray} \label{inv}
\bm{A}^{-1}(\bm{y})=\frac{\bm{H}^\top \bm{y}+\bm{1}_D}{D}.
\end{eqnarray}

Hence $\bm{A}^{-1}(\bm{y})\in\mathbb{S}^{d}$ if and only if every coordinate
of $\bm{A}^{-1}(\bm{y})$ is nonnegative: $\bm{A}^{-1}$ can fail to lie within
the simplex only by producing one or more negative coordinates, i.e.\ by
crossing one of the $D$ hyperplanes $\bm{H}_j=\{\bm{y}:\bm{h}_j^\top \bm{y}=-1\}$, $j=1,\dots,D$, whose intersections with $\mathbb{R}^d$ correspond to the facets $F_j=\{x_j=0\}$ of $\mathbb{S}^d$. This motivates a Gaussian latent--variable model on $\mathbb{R}^d$, censored at these $D$ hyperplanes, as a generative model for $D$--part compositions. We construct the sampler in three steps.

\textbf{Step 1}: For each $j=1,\dots,D$, the event $\bm{A}^{-1}(\bm{y})_j=0$
is equivalent to the scalar condition $\bm{h}_j^\top \bm{y}=-1$, i.e.\ to
$\bm{y}$ lying on the hyperplane $\bm{H}_j$ with unit normal
$\bm{n}_j=\bm{h}_j/\lVert\bm{h}_j\rVert$ and signed distance
$c_j=-1/\lVert\bm{h}_j\rVert$ from the origin. For each $j$ choose an
orthogonal matrix $\bm{B}_j\in\mathbb{R}^{d\times d}$ whose first row is
$\bm{n}_j^\top$ (completing $\bm{n}_j$ to an orthonormal basis, e.g.\ via
the Gram--Schmidt orthonormalization process). Writing $\bm{z}=\bm{B}_j\bm{y}$
for the coordinates of $\bm{y}$ in this rotated basis,
$z_1=\bm{n}_j^\top\bm{y}=\bm{h}_j^\top\bm{y}/\lVert\bm{h}_j\rVert$, so that
\begin{eqnarray*}
\bm{A}^{-1}(\bm{y})_j=0 \iff z_1=c_j .
\end{eqnarray*}
Thus $\bm{H}_j$ is, after rotation, a univariate threshold event on the first coordinate, the role of $(\bm{B}_j,c_j)$ is purely to reduce the $D$--dimensional geometric condition to a one--dimensional one, and $\{(\bm{B}_j,c_j)\}_{j=1}^D$ need only be computed once, prior to sampling.

\textbf{Step 2}: If a draw $\bm{y}$ violates the simplex constraint through
exactly one facet $j$ (i.e.\ $\bm{A}^{-1}(\bm{y})_j<0$ and $\bm{A}^{-1}(\bm{y})_k\ge0$ for all $k\ne j$), we do not discard it but instead censor it onto $\bm{H}_j$: in the rotated coordinates $\bm{z}=\bm{B}_j\bm{y}$, replace the offending coordinate $z_1$ by its boundary value $c_j$, leaving the remaining $d-1$ coordinates, which parametrize the tangent directions of $\bm{H}_j$, unchanged, and rotate
back,
\begin{eqnarray*}
\bm{y}' = \bm{B}_j^\top\big(c_j,\,z_2,\dots,z_d\big)^\top, \qquad
\bm{h}_j^\top\bm{y}'=-1.
\end{eqnarray*}
This is an orthogonal projection of $\bm{y}$ onto $\bm{H}_j$: it is the closest point on $\bm{H}_j$ to $\bm{y}$ in Euclidean distance, and it is exact, so that $\bm{A}^{-1}(\bm{y}')_j=0$. This guarantees only that coordinate $j$ is zeroed out; it does not by itself guarantee that the remaining coordinates of $\bm{A}^{-1}(\bm{y}')$ stay nonnegative. When $\bm{y}$ lies close to an edge or vertex of the simplex (facet $j$ and some other facet $k$ are nearly simultaneously violated), the projection can occasionally overshoot past facet $k$'s boundary as well, so this case must be checked explicitly and, if it occurs, rejected (see \textbf{Step 3}, item 4 below).

\textbf{Step 3}: Fix a sample size $n$ and Gaussian latent parameters
$(\bm{\mu},\bm{\Sigma})$. For $i=1,\dots,n$:
\begin{enumerate}
\item Draw $\bm{y}\sim\mathcal N_d(\bm{\mu},\bm{\Sigma})$ and set
$\bm{x}=\bm{A}^{-1}(\bm{y})$, with violated--facet set
$E=\{j: x_j<0\}$.
\item If $|E|\ge2$ ($\bm{y}$ violates more than one facet simultaneously),
reject and redraw.
\item If $E=\varnothing$, accept $\bm{X}_i=\bm{x}$ as an interior point of
$\mathbb{S}^d$.
\item If $E=\{j\}$, censor onto that facet using Step 2 to obtain
$\bm{y}'$ and $\bm{x}'=\bm{A}^{-1}(\bm{y}')$. If $x'_k<0$ for some
$k\ne j$ (the projection has overshot an adjacent facet), reject and
redraw from Step 3.1; otherwise accept $\bm{X}_i=\bm{x}'$, with $x_j$
set to exactly $0$, and record this observation as a boundary (zero)
value on component $j$.
\end{enumerate}

\subsection{Probability of a zero on the $j$--th face of the simplex}
Once $(\hat{\bm\mu}, \hat{\bm\Sigma})$ have been obtained via the EM
algorithm, the probability of a structural zero on face $j$ admits a closed form,
since it reduces to a univariate normal tail probability. The event $x_j = 0$ corresponds to the hyperplane $\bm{H}_j = \{ \bm{y} : \bm{h}_j^\top \bm{y} = -1 \}$.
Since $\bm{Y} \sim N_d(\bm\mu, \bm\Sigma)$, the projection
$\bm{n}_j^\top \bm Y$ is univariate normal,
\begin{equation*}
\bm{n}_j^\top \bm Y \sim
N\!\left( \bm{n}_j^\top \bm{\mu},\ \bm{n}_j^\top \bm{\Sigma}\bm{n}_j \right),
\end{equation*}
and the face--violation (zero) event is $\{ \bm{n}_j^\top \bm{Y} \lessgtr c_j \}$.
Hence, plugging in the MLE estimates of the parameters, we can estimate the probability by
\begin{equation*}
\widehat{P}(x_j = 0) =
\Phi\!\left(
\frac{c_j - \bm{n}_j^\top \hat{\bm\mu}}
     {\sqrt{\bm{n}_j^\top \hat{\bm\Sigma}\, n_j}}
\right).
\end{equation*}

Since $\lVert \bm{h}_j \rVert = \sqrt{\frac{D-1}{D}}$ for every $j$, we obtain that $c_j = -\sqrt{\frac{D}{D-1}}$, which is the same constant across all $D$ components, because of the symmetry of the Helmert construction. 

Note however that this is the marginal probability of a zero appearance in each component. A better strategy is to estimate the probability via Monte--Carlo. Generate random values from the estimated zero--censored model and then compute the proportion of zero values in each component.

\subsection{Goodness--of--fit test via a simulated $\chi^2$ test}
To assess whether the fitted zero--censored model adequately describes the
observed pattern of structural zeros, we adopted a goodness--of--fit diagnostic similar to the one \citet{butler2008} performed. We test the fit of the estimated model against the observed frequency of zeros in each component using a goodness--of--fit test on $D+1$ mutually exclusive, exhaustive categories: the $n_1$ compositions with no zero (interior points) and, for $j=1,\dots,D$, the compositions with a recorded zero on component $j$. Let $O_0$ denote the observed number of interior compositions and $O_j$ the observed number of compositions with a zero on component $j$, so that $\sum_{j=0}^{D} O_j = n$.

The corresponding cell probabilities are obtained from the fitted model
$(\hat{\bm\mu}, \hat{\bm\Sigma})$ by simulation: a large sample of latent vectors is drawn from $N_d(\hat{\bm\mu}, \hat{\bm\Sigma})$, mapped back to the simplex via $\bm A^{-1}(\bm y) = (\bm H^\top \bm y + \bm 1_D)/D$, and the accept/reject rule of Section~2.8 is applied (draws for which more than one component is negative, or for which the subsequent facet projection would leave another component negative, are discarded and redrawn). The resulting empirical proportions estimate $\hat p_j = \widehat{P}(x_j = 0)$ for $j=1,\dots,D$, and the interior probability is obtained as the complement,
\begin{equation*}
\hat p_0 = 1 - \sum_{j=1}^{D} \hat p_j.
\end{equation*}
The expected count in cell $j$ is then $E_j = n\,\hat p_j$, $j=0,\dots,D$,
with $\sum_{j=0}^{D} E_j = n$ by construction. The observed goodness--of--fit
statistic is the likelihood--ratio test statistic
\begin{equation*}
G^2_{\text{obs}} = 2\sum_{j=0}^{D} O_j \log\!\left(\frac{O_j}{E_j}\right).
\end{equation*}

Since $\hat{\bm\mu}$ and $\hat{\bm\Sigma}$ are estimated from the same data $\bm x$ being tested, comparing $G^2_{\text{obs}}$ to a bootstrap null distribution generated from $(\hat{\bm\mu},\hat{\bm\Sigma})$ alone would understate its true variability. We therefore obtain the null distribution by a nested (double) parametric bootstrap: for $b=1,\dots,B$ replications, a sample $\bm y_b$ of size $n$ is generated from the fitted zero--censored model via the same censored--Gaussian generator used to fit the model; the model is then re--estimated on $\bm y_b$ to obtain $(\hat{\bm\mu}_b, \hat{\bm\Sigma}_b)$ when $\bm y_b$ contains at least one censored observation, or by the ordinary sample mean and covariance of the $\alpha$--transformed data when $\bm y_b$ happens to contain none, since the model then reduces to an uncensored multivariate normal. The resulting counts $O_j^{(b)}$ are tabulated over the same $D+1$ categories, expected counts $E_j^{(b)} = n\,\hat p_j^{(b)}$ are recomputed from $(\hat{\bm\mu}_b,\hat{\bm\Sigma}_b)$ by the same simulation procedure,
and
\begin{equation*}
G^2_{(b)} = 2\sum_{j=0}^{D} O_j^{(b)} \log\!\left(\frac{O_j^{(b)}}{E_j^{(b)}}\right)
\end{equation*}
is computed against the replicate--specific expected counts. The simulated $p-value$ is then
\begin{equation*}
p-value = \frac{1+\sum_{b=1}^{B} \mathbb{I}\!\left\{ G^2_{(b)} \ge G^2_{\text{obs}} \right\}}{1+B},
\end{equation*}
which does not rely on the large--sample $\chi^2$ approximation, remains
valid under the sparse--cell regime typical of structural--zero data, and
correctly accounts for the parameters having been estimated from the data
under test.

\section{Simulations} \label{sec:sims}
\subsection{An example of the zero--censored model on $\mathbb{S}^2$}
\begin{figure}[!ht]
\centering
\begin{tabular}{cc}
\includegraphics[scale = 0.65]{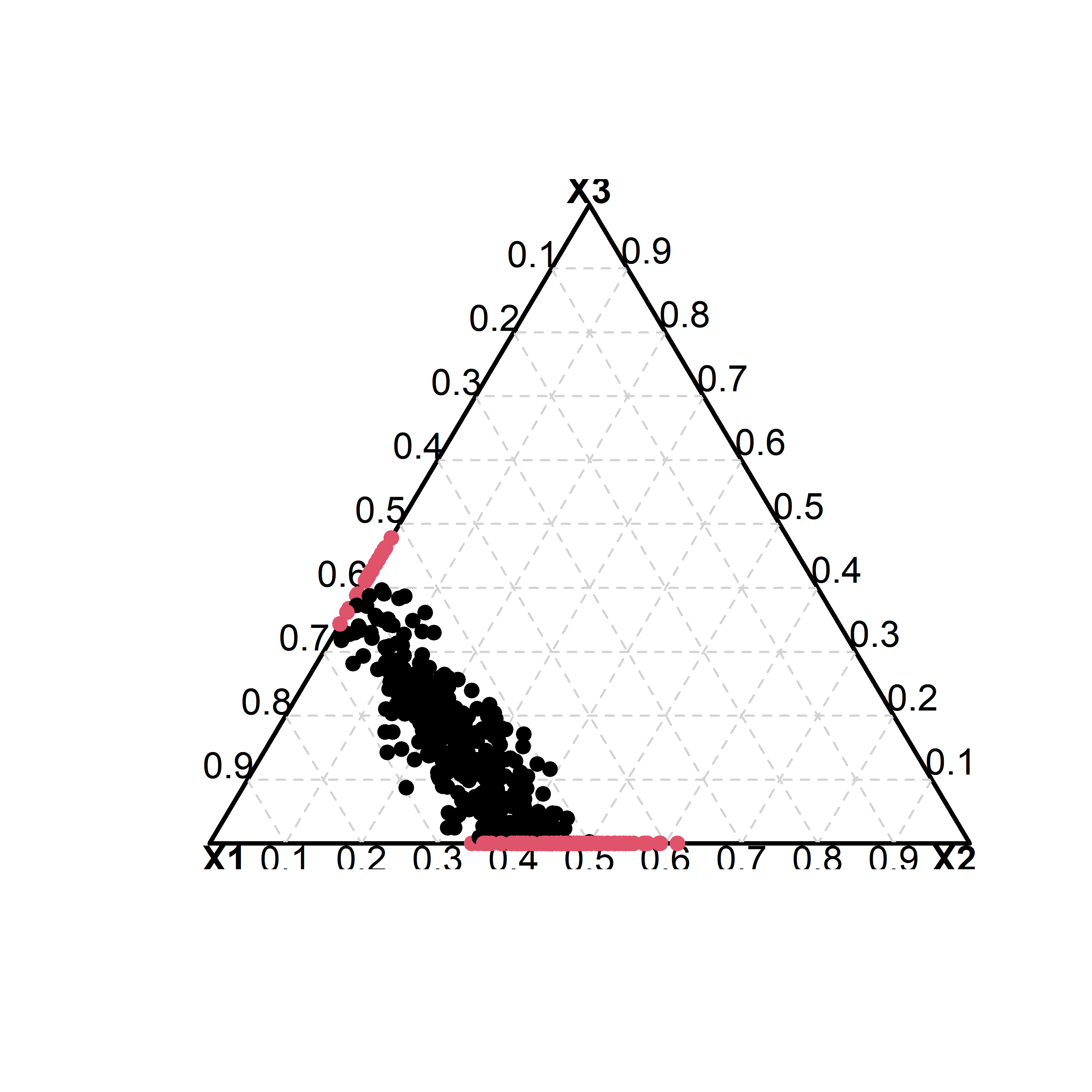} &
\includegraphics[scale = 0.65]{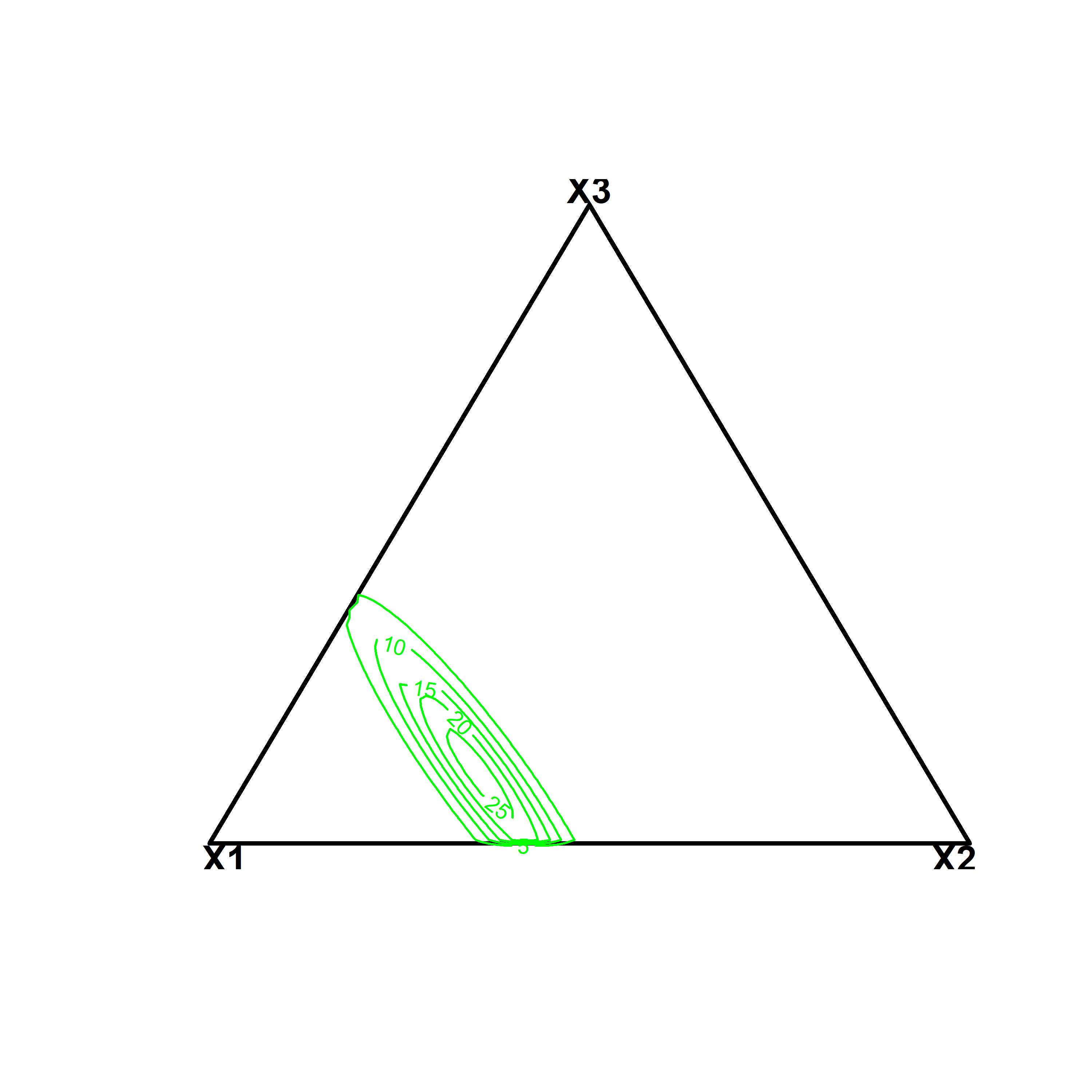} \\
(a) Ternary plot of the data. & (b) Contour plot of the fitted zero--censored model.
\end{tabular}
\caption{On the left is the ternary diagram where the red colour indicates the points which lie on the boundaries of the simplex. The contour plots of the zero--censored model based on the estimated parameters are shown on the right.} 
\label{zerocensor}
\end{figure}

Figure \ref{zerocensor} shows a simulated example of the zero--censored model. Data of size $500$ were generated from the following multivariate normal
\begin{eqnarray*}   
N_2\left(
\begin{array}{cc}
(0.625 , & 0.821)
\end{array},
\left( \begin{array}{cc}
 0.149 & -0.200 \\
-0.200 & 1.523 
\end{array} \right)
\right).
\end{eqnarray*}
There were $197$ vectors with one zero value. We then fitted the zero--censored model to the data by maximizing the log--likelihood (\ref{zerolog3}), and the estimated parameters of this normal distribution, used in this random vector generation, were
\begin{eqnarray*}
\hat{\bm{\mu}}=\left(0.656, 0.788 \right) \ \ \text{and} \ \ 
\hat{\bm{\Sigma}}=\left( \begin{array}{cc} 0.129 & -0.132  \\  -0.132 & 1.477  \end{array} \right).
\end{eqnarray*}
Figure \ref{zerocensor} shows the ternary plot of the data along with the contours of the zero--censored model computed from the estimated parameters. 

\subsection{Performance of the MLE}
We then performed a small scale simulation study to see the effect of the proportion of zero values in the MLE. We generated data of sample sizes $n=(100,200,\ldots,1000)$ from a $N_2\left(\bm{\mu},k\bm{\Sigma}\right)$, where $\bm{\mu}$ and $\bm{\Sigma}$ are the mean vector and covariance matrix, respectively, of the previous example and $k=1,2,\ldots,5$. As $k$ increases, so do the number of zero values. We fit the model in the simulated data and compute the discrepancy between the observed and the estimated mean vector via the Euclidean distance, and the discrepancy between the true and the estimated covariance matrix using the metric of \cite{forstner2003}. We then aggregated the discrepancies over 500 repetitions. 

\begin{figure}[!ht]
\centering
\begin{tabular}{cc}
\includegraphics[scale = 0.45]{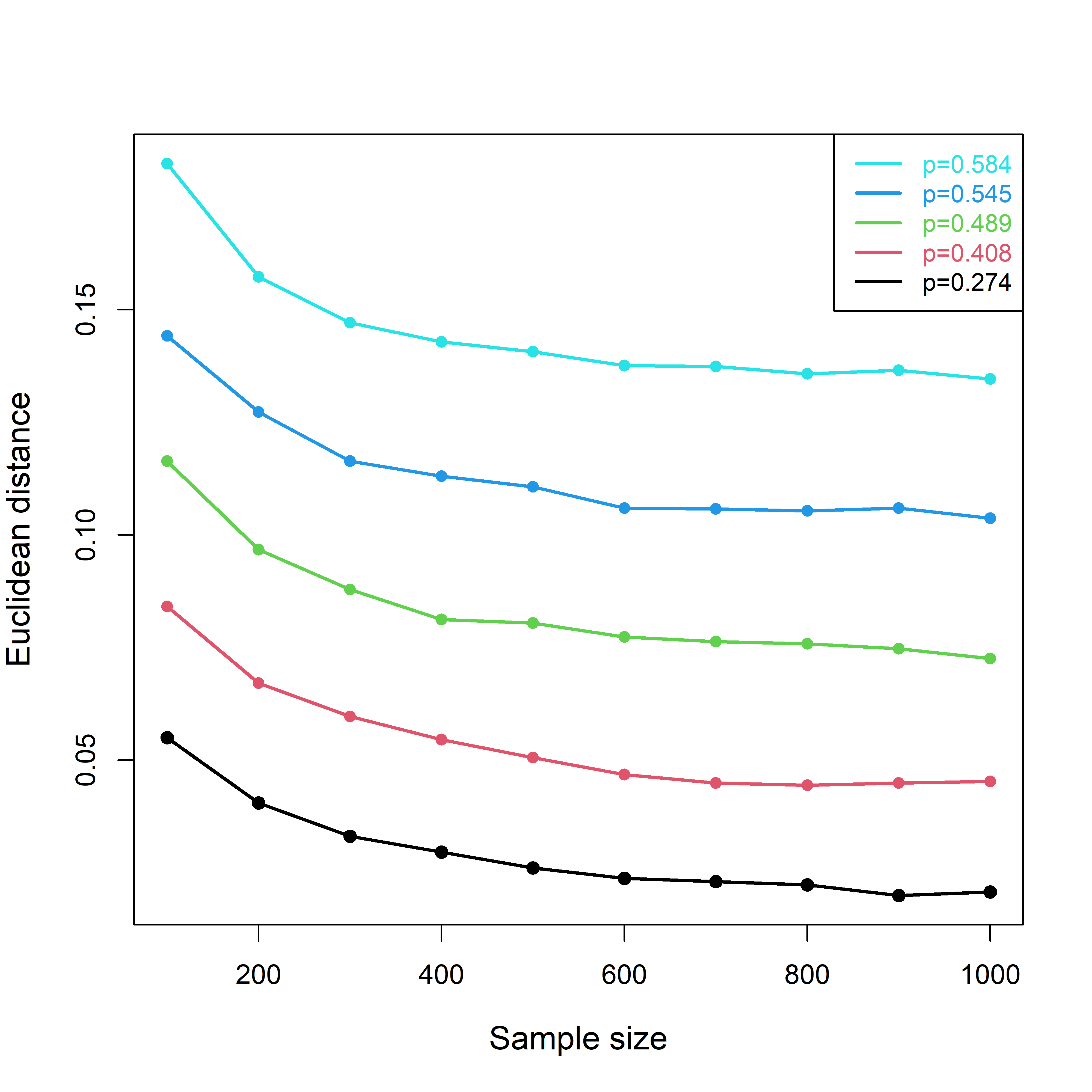} &
\includegraphics[scale = 0.45]{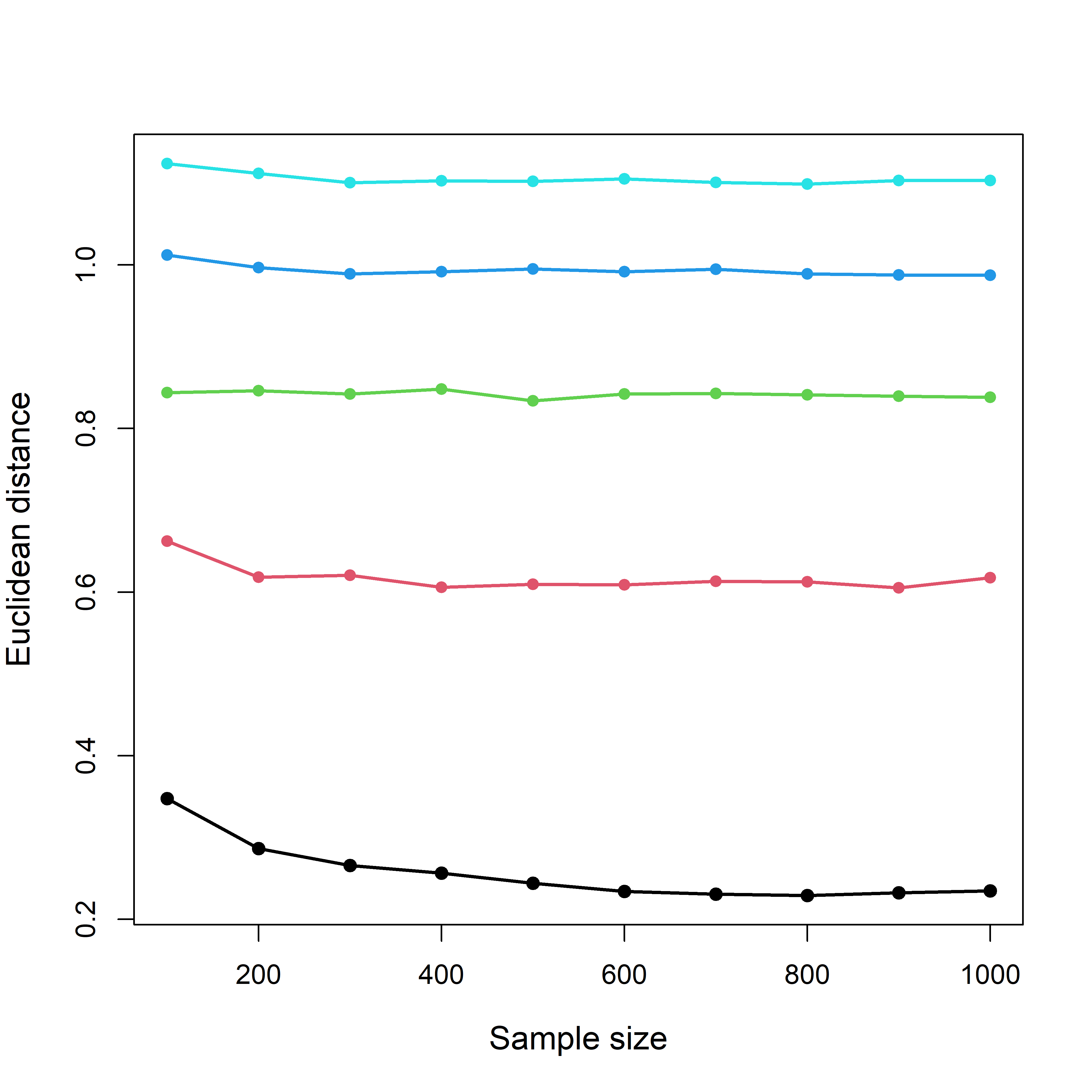} \\
(a) Euclidean distances for mean vector. & (b) Covariance distances.
\end{tabular}
\caption{On the left is the Euclidean distance between the true and the estimated mean vector, and on the right is the distance between the true and the estimated covariance matrix, for each sample size. The different colours indicate the probability of having a structural zero in one component of the simplex.} 
\label{dists}
\end{figure}

Figure \ref{dists} presents the results. Evidently, as the number of censored events (number of zero values) increases, so does the distance between the true and estimated parameters. The discrepancy in the mean vector decreases, with increasing sample sizes, but the discrepancy in the covariance matrix decreases at a slower rate, until both reach a plateau.   

\subsection{Computational cost}
\begin{figure}[!h]
\includegraphics[scale=0.5,trim=0 0 0 0]{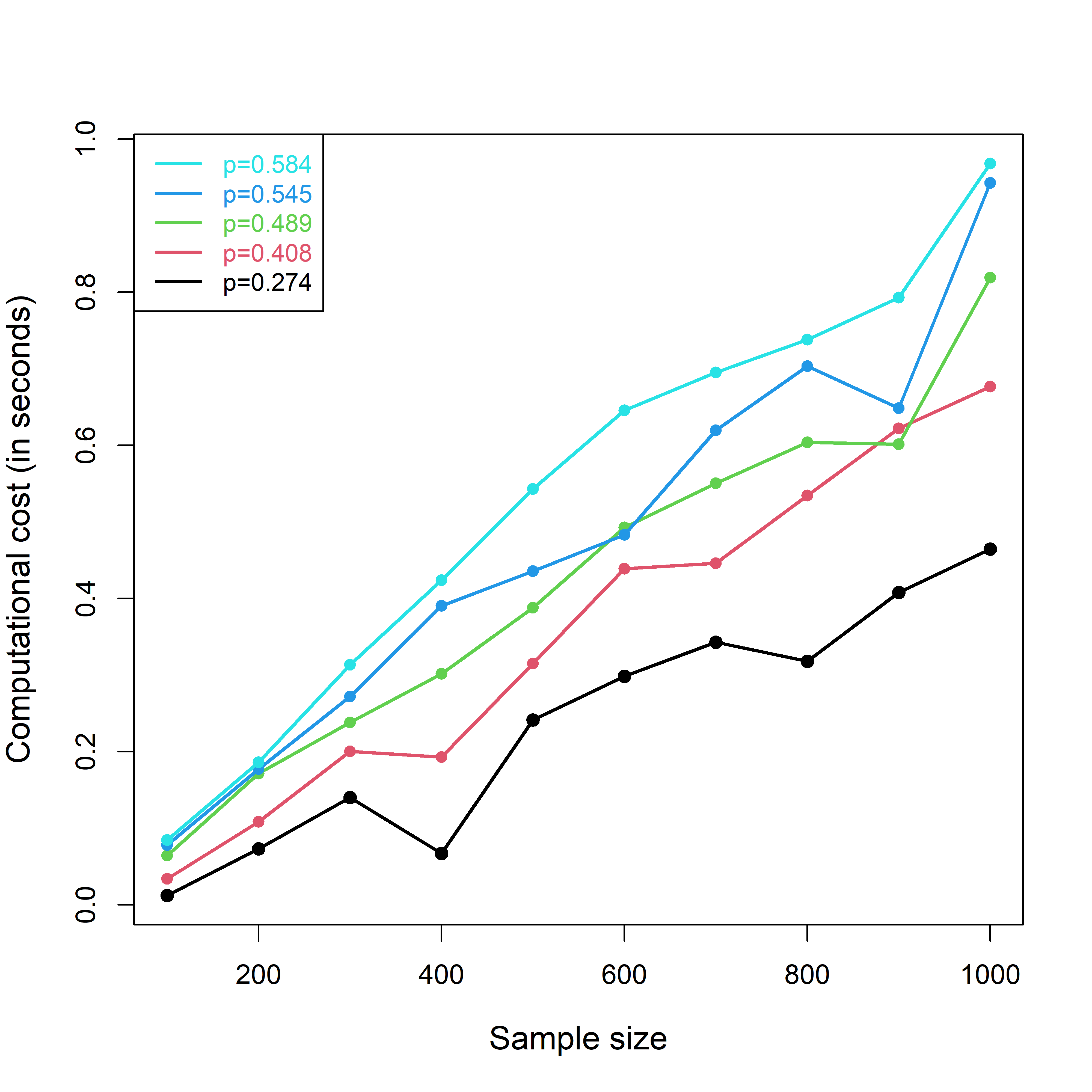}
\centering
\caption{Computational cost of the MLE process as a function of the sample size. The different colours indicate the probability of having a structural zero in one component of the simplex.} 
\label{times}
\end{figure}

We also computed the running time (computational cost) to fit the zero--censored model in the above simulations. Using the same parameters and sample sizes we computed the computational cost repeating the MLE process 20 times and aggregated the results. Figure \ref{times} contains the running times (in seconds). 

Evidently, as the number of censored events (number of zero values) increases, so does the computational cost. Further, we can observe that the computational cost increases linearly in the sample size. 

\section{Real data analysis} \label{sec:real}
We will illustrate the performance of the zero--censored model using three real datasets. 

\subsection{Foraminiferal data}
\begin{figure}[!h]
\includegraphics[scale=0.45,trim=0 0 0 0]{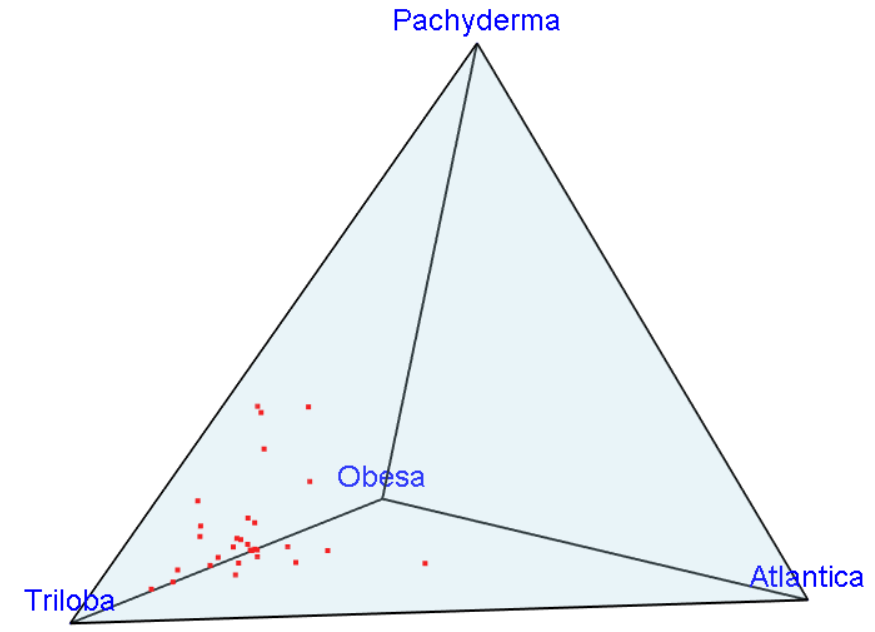}
\centering
\caption{Pyramid diagram of the foraminineral data.} 
\label{data34}
\end{figure}

This first dataset consists of foraminiferal (marine plankton species) compositions measured at different depths \citep{ait2003}. The data consist of 4 components, Triloba, Obesa, Pachyderma and Atlantica. There are $5$ compositional vectors having a zero value, either in the third or the fourth component. Since there are $4$ components we can plot them on a pyramid in Figure \ref{data34}. 

The estimated parameters are
\begin{eqnarray*}
\hat{\bm{\mu}}=\left(1.096, 1.226, 0.978\right) \ \ \text{and} \ \
\hat{\bm{\Sigma}}=\left[
\begin{array}{rrr}
0.352  & -0.034 & 0.040 \\
-0.034 &  0.184 & -0.013 \\
0.040  & -0.013 & 0.082 \\
\end{array} \right].
\end{eqnarray*}

The goodness--of--fit test produced a $p-value$ equal to 0.001, indicating that the model has not captured the observed pattern of zeros satisfactorily. This means that the marginal frequency of "zero on $j$--th component" does not match the observed pattern of zeros.

\subsection{Glass data}
The second dataset consists of 27 glasses with 11 chemical components analyzed twice and comes from archaeometry \citep{baxter2006b}. Out of the 54 compositional vectors, 20 contain one zero value. 

The estimated parameters are
\begin{eqnarray*}
\hat{\bm{\mu}}=\left(-0.020, -1.649, -5.394, 1.953, 1.748, 1.484, 1.295, 0.913, 1.040 \right) \ \ \text{and} \\
\\
\hat{\bm{\Sigma}}=10^{-2}\left[
\begin{array}{rrrrrrrrr}
0.034 & 0.162 & -0.017 & -0.046 & -0.036 & -0.026 & -0.025 & -0.194 & 0.007 \\ 
0.162 & 3.935 & -5.134 & -0.076 & -0.104 & -0.048 & -0.040 & -0.501 & 0.101 \\ 
-0.017 & -5.134 & 9.332 & -0.566 & -0.248 & -0.221 & -0.237 & -1.240 & -0.107 \\ 
-0.046 & -0.076 & -0.566 & 0.224 & 0.103 & 0.079 & 0.081 & 0.520 & 0.000 \\ 
-0.036 & -0.104 & -0.248 & 0.103 & 0.069 & 0.051 & 0.051 & 0.350 & 0.005 \\ 
-0.026 & -0.048 & -0.221 & 0.079 & 0.051 & 0.041 & 0.040 & 0.276 & 0.006 \\ 
-0.025 & -0.040 & -0.237 & 0.081 & 0.051 & 0.040 & 0.042 & 0.265 & 0.005 \\ 
-0.194 & -0.501 & -1.240 & 0.520 & 0.350 & 0.276 & 0.265 & 2.020 & -0.038 \\ 
0.007 & 0.101 & -0.107 & 0.000 & 0.005 & 0.006 & 0.005 & -0.038 & 0.082 \\ 
\end{array} \right].
\end{eqnarray*}

The goodness--of--fit test produced a $p-value$ equal to 0.116, indicating a satisfactory fit of the observed pattern of zeros.

\subsection{Time budget data}
The time budget dataset \citep{hardle2007} carries information on $28$ individuals, for each person the time allocation in $10$ activities is known. The interest lies in the relative amount of time each person spent on $10$ categories of activities over $100$ days (the total is $100 \times 24=2400$ hours fixed for every row) in $1976$. The special feature of these data is that they contain some zero values. Some activities have zero allocation, for instance one woman did not spend even an hour on transportation linked to professional activity and four women did not spend any hour on occupation linked to children. This means that we have five compositions which have one zero in one component only.  

The estimated parameters are
\begin{eqnarray*}
\hat{\bm{\mu}}=\left(1.075, -0.030,  0.860,  0.383,  0.367,  0.222, -2.417,  0.568, -0.465 \right) \ \ \text{and} \\
\\
\hat{\bm{\Sigma}}=\left[
\begin{array}{rrrrrrrrr}
0.289 &  0.574 & 0.109 & 0.076 & 0.019 & 0.035 & 0.038 &  0.008 & 0.030 \\
0.574 &  1.240 & 0.200 & 0.119 & 0.020 & 0.063 & 0.069 & -0.020 & 0.002 \\
0.109 &  0.200 & 0.050 & 0.035 & 0.013 & 0.020 & 0.019 &  0.015 & 0.019 \\
0.076 &  0.119 & 0.035 & 0.037 & 0.013 & 0.009 & 0.013 &  0.015 & 0.027 \\
0.019 &  0.020 & 0.013 & 0.013 & 0.008 & 0.007 & 0.008 &  0.010 & 0.014 \\
0.035 &  0.063 & 0.020 & 0.009 & 0.007 & 0.020 & 0.018 &  0.010 & 0.010 \\
0.038 &  0.069 & 0.019 & 0.013 & 0.008 & 0.018 & 0.021 &  0.005 & 0.019 \\
0.008 & -0.020 & 0.015 & 0.015 & 0.010 & 0.010 & 0.005 &  0.029 & 0.003 \\
0.030 &  0.002 & 0.019 & 0.027 & 0.014 & 0.010 & 0.019 &  0.003 & 0.079 \\
\end{array} \right].
\end{eqnarray*}

We then computed the probability of a zero value in each of the 10 faces of the simplex and multiplied these probabilities by 28 to obtain the estimated number of zeros. Table \ref{zerotable} contains the observed and the estimated number of zeros for the time budget dataset. The goodness--of--fit test produced a $p-value$ equal to 0.635, indicating a satisfactory fit of the observed pattern of zeros. 

\begin{small} 
\begin{table}[!h] 
\caption{Time budget data: observed and estimated number of zeros for every component.}
\label{zerotable}
\begin{center}
\begin{tabular}{l|cccccccccc} \toprule
Components      & $X_1$  & $X_2$ & $X_3$ & $X_4$ & $X_5$ & $X_6$ & $X_7$ & $X_8$ & $X_9$ & $X_{10}$  \\ \midrule
Observed        &    &   &    &   &   &   &   &   &   &   \\ 
number of zeros & 0  & 1 & 0  & 4 & 0 & 0 & 0 & 0 & 0 & 0 \\ \midrule
Estimated       &    &   &    &   &   &   &   &   &   &   \\     
number of zeros & 0.61 & 0.98 & 2.20 & 3.97 & 0.01 & 0.00 & 0.00 & 0.00 & 0.15 & 0.00  \\ \bottomrule
\end{tabular}
\end{center}
\end{table}
\end{small}

\section{Conclusions}
We developed a parametric zero--censored model for compositional data with zero values. The use of another multivariate model, such as the multivariate skew normal distribution \citep{azzalini1996} could also be utilized but the difficulty with this distribution is that more parameters need to be estimated, thus making the estimation procedure more difficult. This of course does not exclude the possibility of using this model. The advantage of the proposed model is that it can be fitted computationally efficiently via the EM algorithm. The disadvantage though is that it allows only one zero per compositional vector. 

The simulation studies revealed that the model is harder to fit as the number of components with one zero value increases, and that it is computationally efficient to fit. The illustration of the performance of the zero--censored model on the real data revealed that the model does not always fit the observed pattern of zeros satisfactorily. 

We have created the \textit{R} package \textit{Compositionalzerocens} \citep{compositionalzerocens2026} to fit the zero--censored model, draw random values, estimate the probability of zero values in one component, perform the goodness--of--fit test, available to download from \href{https://cran.r-project.org/web/packages/Compositionalzerocens/index.html}{CRAN}. Future research is directed to generalizing the model to allow for more than one zero values in each compositional vector. To do so, we need to modify the model along similar lines to \cite{leininger2013}. \\
\\
\textbf{Funding statement}: No funding was received. \\
\textbf{COI}: The authors declare no conflict of interest.



\bibliographystyle{apalike}
\bibliography{vivlio}

\end{document}